\documentclass{article}
\usepackage{graphicx}
\usepackage[english]{babel}

\title{
\includegraphics[width=0.35\textwidth]{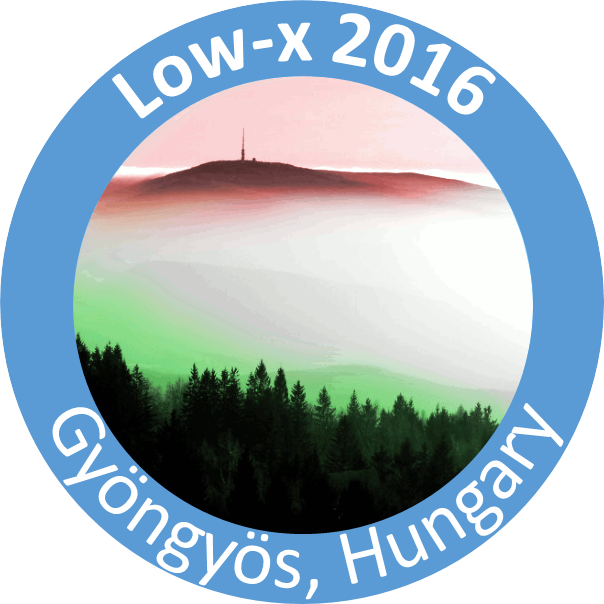}\\[1cm]
Diphoton production in lead-lead \\and proton-proton UPC}
\author{{M. K{\l}usek-Gawenda$^1$, A. Szczurek$^1$\footnote{Also at University of Rzesz{\'o}w, PL-35-959 Rzesz{\'o}w, Poland.},}\\[1ex]
$^1$Institute of Nuclear Physics Polish Academy of Sciences,\\
Radzikowskiego 152, PL-31-342 Krak{\'o}w, Poland\\
}

\begin{document}

\fontfamily{lmss}\selectfont
\maketitle

\begin{abstract}
We discuss diphoton semi(exclusive) production in ultraperipheral $PbPb$ collisions 
at energy of $\sqrt{s_{NN}}=$ 5.5 TeV (LHC). The nuclear calculations are based on equivalent photon approximation in the impact parameter space. 
The cross sections for elementary $\gamma \gamma \to \gamma \gamma$ subprocess are calculated including three different mechanisms: 
box diagrams with leptons and quarks in the loops, a VDM-Regge contribution with virtual 
intermediate hadronic excitations of the photons and 
the two-gluon exchange contribution (formally three-loops) to elastic photon-photon
scattering in the high-energy approximation. 
We got relatively high cross sections in $PbPb$ collisions ($306$ nb). 
This opens a possibility to study the $\gamma \gamma \to \gamma \gamma$ (quasi)elastic scattering at the LHC. 
We find that the cross section for elastic $\gamma\gamma$ scattering could be measured in the lead-lead collisions
for the diphoton invariant mass up to $W_{\gamma\gamma} \approx 15-20$ GeV.
We identify region(s) of phase space
where the two-gluon exchange contribution becomes important ingredient compared to box and
nonperturbative VDM-Regge mechanisms.
We perform a similar analysis for the $pp \to pp\gamma\gamma$ reaction at energy of
$\sqrt{s_{NN}}=7$ and $100$ TeV.
\end{abstract}

\section{Introduction}

In classical Maxwell theory photons/waves/wave packets do not interact.
In contrast, in quantal theory they can interact via quantal fluctuations.
So far only inelastic processes, i.e. production of hadrons or jets
via photon-photon fusion could be measured e.g. in $e^+ e^-$ collisions
or in ultraperipheral collisions (UPC) of heavy-ions.
It was realized only recently that ultraperipheral heavy-ions 
collisions can be also a good place where photon-photon elastic
scattering could be tested experimentally \cite{d'Enterria:2013yra,KGLS2016}.

\section{Theory}
\subsection{Elementary cross section}

\begin{figure}[!h]
\centering
\includegraphics[scale=0.25]{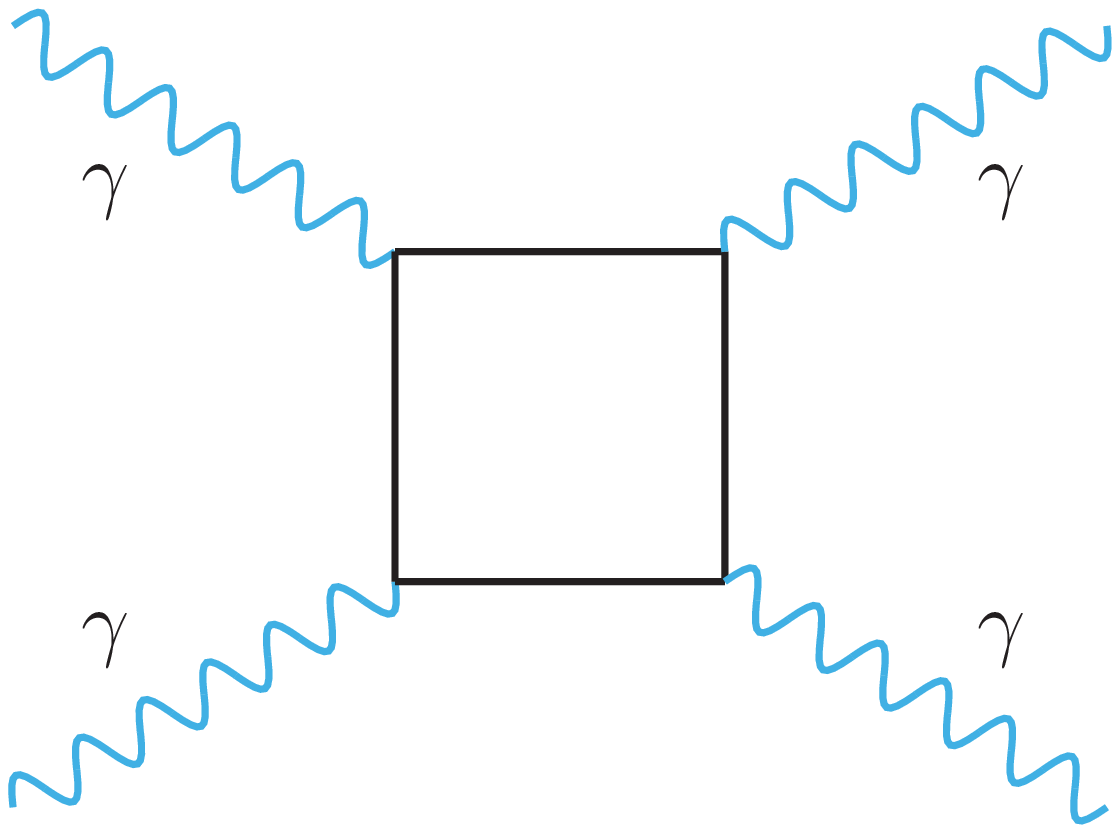}
\includegraphics[scale=0.25]{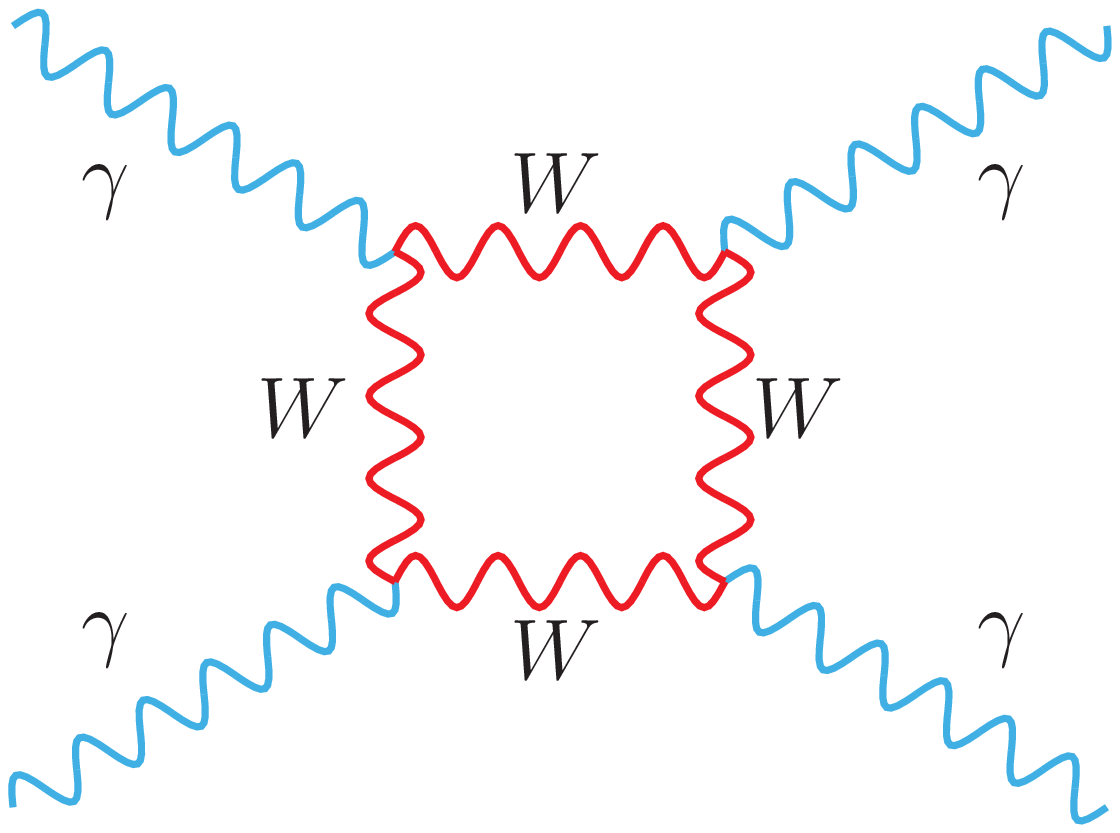}
\includegraphics[scale=0.35]{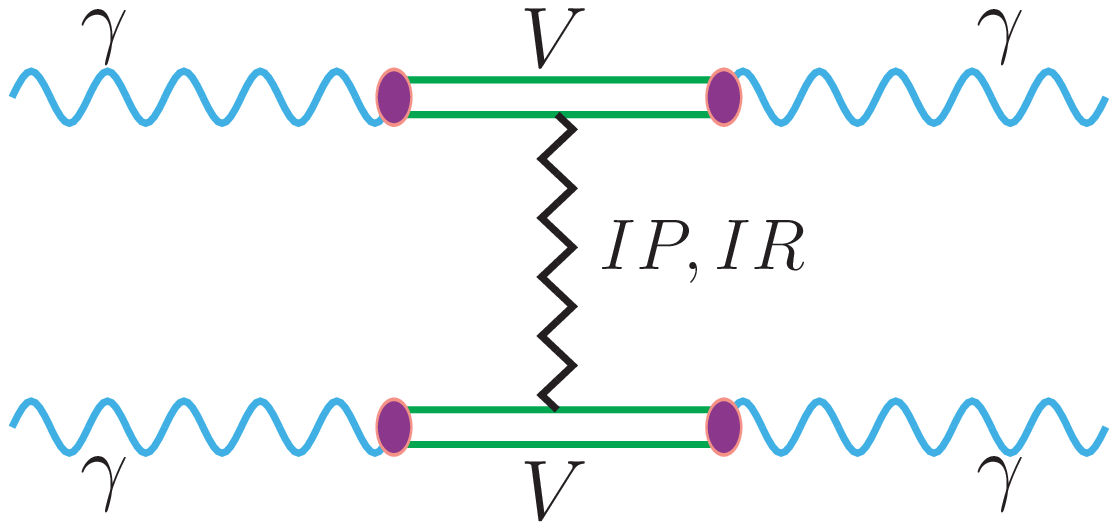}
\includegraphics[scale=0.35]{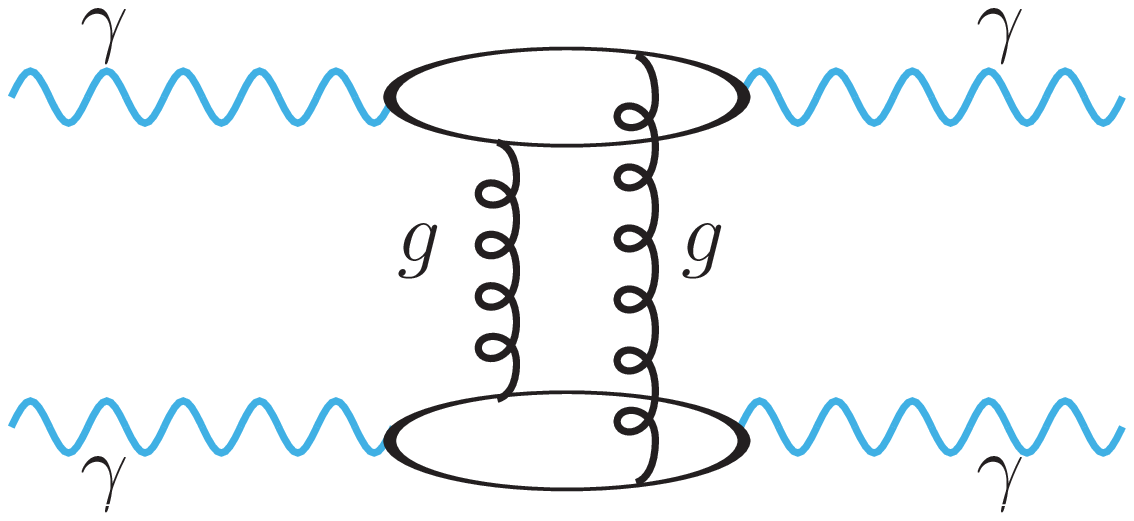}
\caption{Light-by-light scattering mechanisms with 
the lepton and quark loops (first diagram) 
and for the intermediate $W$-boson loop (second diagram). 
Third diagram represents VDM-Regge mechanism 
and the last diagram is for two-gluon exchange.}
\label{fig:diagrams_elementary}
\end{figure}

One of the main ingredients of the formula for calculation of the nuclear cross section
is elementary $\gamma\gamma\to\gamma\gamma$ cross section. 
The lowest order QED mechanisms with elementary particles are shown
in two first diagrams of Fig.~\ref{fig:diagrams_elementary}. 
The first diagram is for lepton and quark loops and it dominates
at lower photon-photon energies ($W_{\gamma\gamma}<2m_W$) 
while the next diagram is for the $W$ (spin-1) boson loops and it becomes 
dominant at higher photon-photon energies (\cite{Bardin2009,Lebiedowicz2013}). 
The one-loop box amplitudes were calculated by using
the Mathematica package {\tt{FormCalc}} and the {\tt{LoopTools}}
library.
We have obtained good agreement when confronting our result with those in 
\cite{Bardin2009,Jikia1993,Bern2001}.
Including higher-order contributions seems to be interesting.
In Ref.~\cite{Bern2001} the authors considered both 
the QCD and QED corrections (two-loop Feynman diagrams)
to the one-loop fermionic contributions in the ultrarelativistic limit
($\hat{s},|\hat{t}|,|\hat{u}| \gg m_f^2$). 
The corrections are quite small numerically so the leading order computations 
considered by us are satisfactory.
In the last two diagrams of Fig.~\ref{fig:diagrams_elementary} we show processes 
that are the same order in $\alpha_{em}$ but higher order in $\alpha_s$. 
Third diagram presents situation where
both photons fluctuate into virtual vector mesons
(here we include three different light vector mesons: $\rho, \omega, \phi$). 
The last diagram shows two-gluon exchange mechanism which is formally three-loop type.
Its contribution to the elastic scattering of photons at high energies 
has been first considered in the pioneering work \cite{Gin87}. 
Indeed in the limit where the Mandelstam variables of 
the $\gamma \gamma \to \gamma \gamma$ process satisfy
$\hat s \gg -\hat t$, $-\hat u$, major simplifications occur and the three-loop process
becomes tractable. 
This corresponds to a near-forward, small-angle, scattering of photons
In our treatment, we go beyond the early work \cite{Gin87} by including 
finite fermion masses, as well as the full momentum structure in the loops, 
and we consider all helicity amplitudes \cite{KGSSz2016}.


\subsection{Nuclear cross section}

\begin{figure}[!h]
\centering
\includegraphics[scale=0.25]{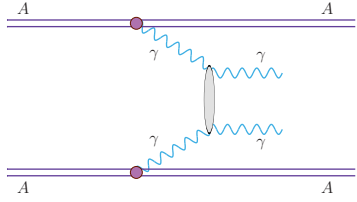}
\caption{Diphoton production in ultrarelativistic UPC of heavy ions.}
\label{fig:diagram_AA_AAgamgam}
\end{figure}
%

The general situation for the $AA \to AA \gamma \gamma$ reaction 
is sketched in Fig.~\ref{fig:diagram_AA_AAgamgam}.
In our equivalent photon approximation (EPA) in the impact parameter space, 
the total (phase space integrated) cross section 
is expressed through the five-fold integral
(for more details see e.g.~\cite{KG2010})
\begin{eqnarray}
\sigma_{A_1 A_2 \to A_1 A_2 \gamma \gamma}\left(\sqrt{s_{A_1A_2}} \right) &=&
\int \sigma_{\gamma \gamma \to \gamma \gamma} 
\left(W_{\gamma\gamma} \right)
N\left(\omega_1, {\bf b_1} \right)
N\left(\omega_2, {\bf b_2} \right) \nonumber \\
&\times & S_{abs}^2\left({\bf b}\right)
2\pi b \mathrm{d} b \, \mathrm{d}\overline{b}_x \, \mathrm{d}\overline{b}_y \, 
\frac{W_{\gamma\gamma}}{2}
\mathrm{d} W_{\gamma\gamma} \, \mathrm{d} Y_{\gamma \gamma} \;,
\label{eq:EPA_sigma_final_5int}
\end{eqnarray}
where $N(\omega_i,{\bf b_i})$ are photon fluxes,
$W_{\gamma\gamma}=\sqrt{4\omega_1\omega_2}$
and $Y_{\gamma \gamma}=\left( y_{\gamma_1} + y_{\gamma_2} \right)/2$ 
is a invariant mass and a rapidity of the outgoing $\gamma \gamma$ system. 
Energy of photons is expressed through $\omega_{1/2} = W_{\gamma\gamma}/2 \exp(\pm Y_{\gamma\gamma})$.
$\bf b_1$ and $\bf b_2$ are impact parameters 
of the photon-photon collision point with respect to parent
nuclei 1 and 2, respectively, 
and ${\bf b} = {\bf b_1} - {\bf b_2}$ is the standard impact parameter 
for the $A_1 A_2$ collision.

The photon flux ($N(\omega,b)$) is expressed through a nuclear charge
form factor. In our calculations we use two different types of the form factor.
The first one, called here realistic form factor, is the Fourier transform of
the charge distribution in the nucleus and the second one is a monopole form factor
which leads to simpler analytical results.
More details can be found e.g. in \cite{KGLS2016,KG2010}.

\subsection{$pp \to pp \gamma\gamma$ process}

We can study the mechanism of elastic photon-photon scattering also
in $pp \to pp \gamma\gamma$ reaction. 
In our calculations we neglect the gap survival factor.
Then the cross section of $\gamma\gamma$ production in proton-proton collisions 
takes the simple parton model form

\begin{equation}
\frac{\mathrm{d}\sigma}{\mathrm{d}y_1 \mathrm{d}y_2 \mathrm{d}^2p_t} = \frac{1}{16\pi^2 \hat{s}^2}
x_1 \gamma^{(el)} (x_1)
x_2 \gamma^{(el)} (x_2) 
\overline{\left|{M}_{\gamma\gamma \to \gamma \gamma}\right|^2} \;.
\label{eq:pp}
\end{equation}
Here $y_{1/2}$ is the rapidity of final state photon, 
$p_t$ is the photon transverse momentum
and $x_{1/2} = p_t/\sqrt{s} \left( \exp\left( \pm y_1 \right) + \exp\left( \pm y_2 \right) \right)$.
In the numerical calculations for the elastic fluxes we shall use a practical parametrization
of Ref. \cite{Dress}. Detailed description of
Eq.~\ref{eq:pp} one can be found in our paper \cite{KGSSz2016}.
 
\section{Results}

\begin{figure}[!h]
\centering
\includegraphics[scale=0.28]{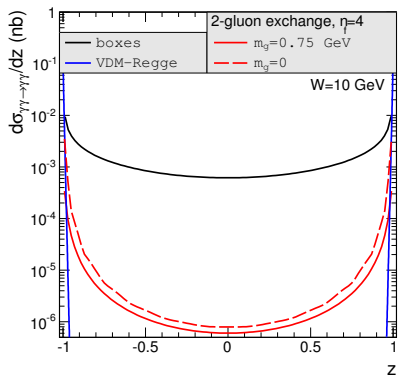}
\caption{Statement of the three considered processes for $W = 10$ GeV.}
\label{fig:dsig_dz_10GeV}
\end{figure}
%

In Fig. \ref{fig:dsig_dz_10GeV} we show contributions of mechanisms 
presented in Fig.~\ref{fig:diagrams_elementary} for fixed value of energy $W=10$ GeV.
The differential cross section is shown as a function of $z = \cos \theta$, 
where $\theta$ is the scattering angle in the $\gamma\gamma$ cms. 
The contribution of the VDM-Regge is concentrated at $z \approx \pm 1$. 
In contrast, the box contribution extends over a broad range of $z$. 
The two-gluon exchange contribution occupies intermediate regions of $z$. 
We need to add though, that the approximations made in the calculation of the two-gluon exchange are justified in a small angle region only. At small $z$ the error can
easily be 100$\%$.
In addition we show the difference between results when
we include gluon mass ($m_g = 750$ MeV - solid line) and for massless particle 
($m_g = 0$ - dashed line).

\begin{figure}[!h]  
\center
\includegraphics[scale=0.28]{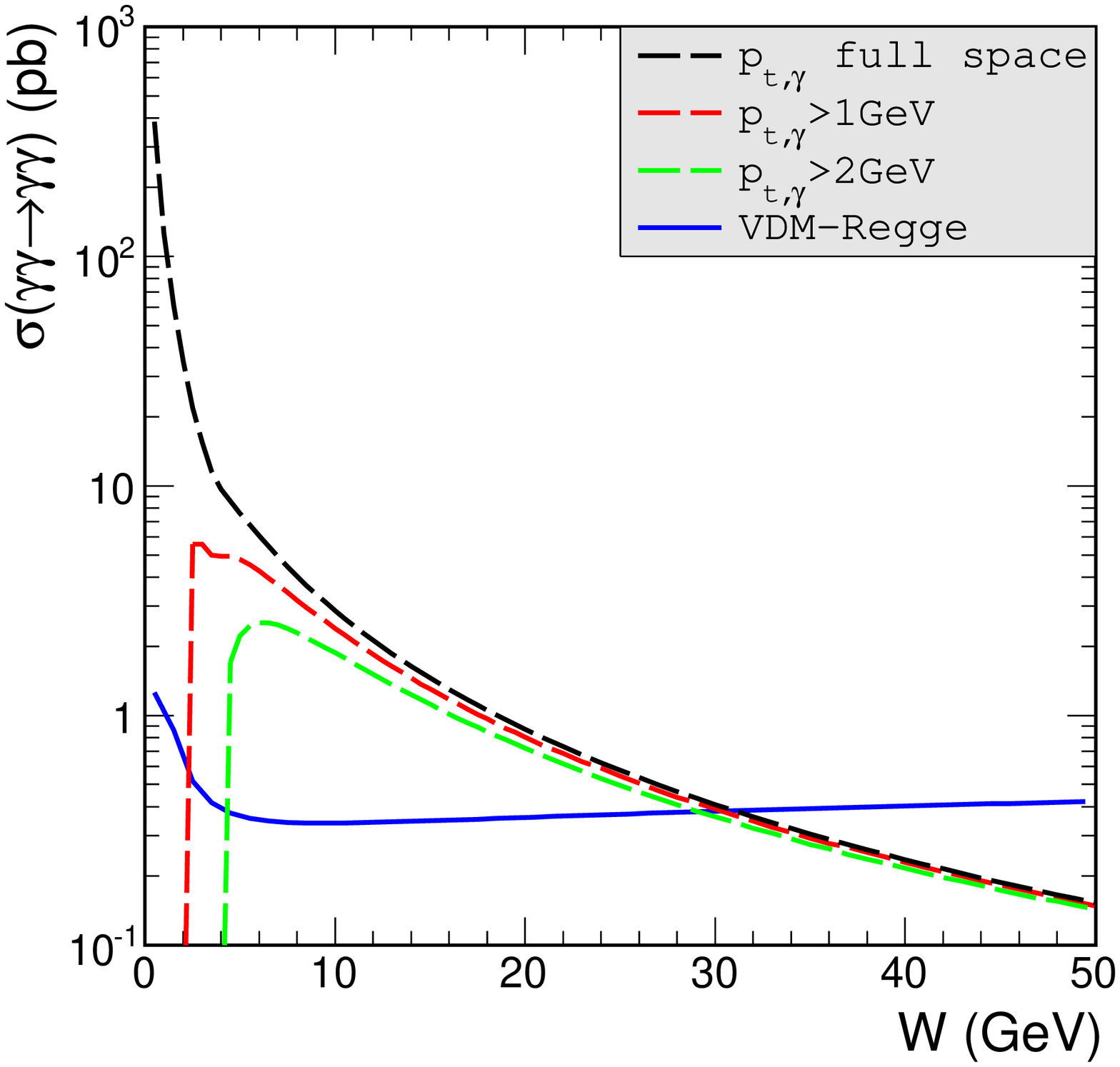}
\includegraphics[scale=0.28]{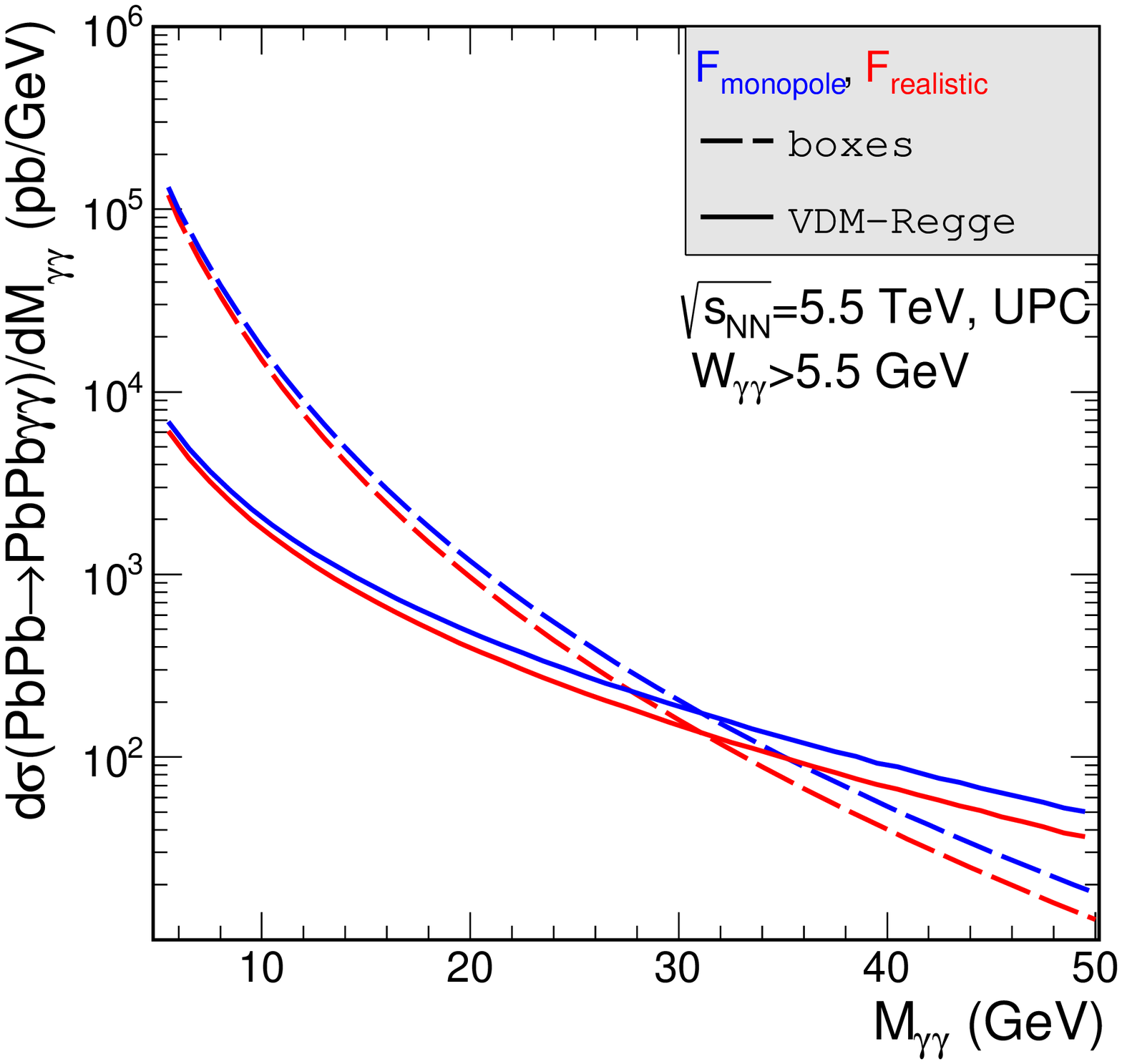}
  \caption{
  Elementary and nuclear cross section for light-by-light scattering.
  The dashed lines show the results for the case
  when only box contributions (fermion loops) are included and 
  the solid lines show the results for the VDM-Regge mechanism.
  Left panel: integrated elementary $\gamma \gamma \to \gamma \gamma$ cross section
  as a function of the subsytem energy. 
  Right panel: differential nuclear cross section as a function of $\gamma\gamma$ invariant mass
  at $\sqrt{s_{NN}}=5.5$ TeV. 
  The distributions with the realistic charge density are depicted by
  the red (lower) lines and the distributions which are calculated
  using the monopole form factor are shown by the blue (upper) lines.
  }
\label{fig:dsig_dw}
\end{figure}

The elementary angle-integrated cross section for the box and 
VDM-Regge contributions is shown in the first panel of Fig.~\ref{fig:dsig_dw}
as a function of the photon-photon subsystem energy.
Lepton and quark amplitudes interfere enhancing the cross section.
For instance in the 4 GeV $<W<$ 50 GeV region, 
neglecting interference effects, the lepton contribution 
to the box cross section is by a factor $5$ bigger than the quark contribution.
Interference effects are large and cannot be neglected.
At energies $W >30$~GeV the VDM-Regge cross section becomes larger
than that for the box diagrams. The right panel of Fig.~\ref{fig:dsig_dw}
shows results for nuclear collisions for the case of realistic charge density (red lines) 
and monopole form factor (blue lines). 
The difference between the results becomes larger with larger values of 
the kinematical variables. The cross section obtained with the monopole 
form factor is somewhat larger.

\begin{figure}[!h]  
\center
\includegraphics[scale=0.28]{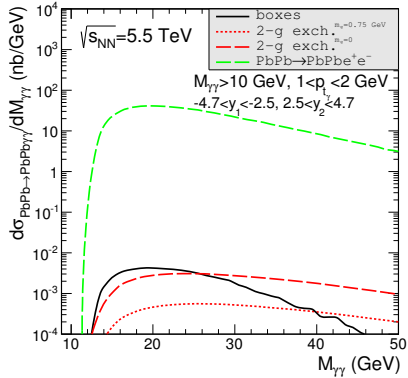}
  \caption{Distribution in invariant mass of photons 
for $M_{\gamma\gamma} > 10$ GeV, $1$ GeV $< p_t < 2$ GeV and $- 4.7 < y_1 < - 2.5$, 
$2.5 < y_2 < 4.7$. In addition, we show (top dashed, green line) a similar distribution 
for $AA \to AA e^+ e^−$.}
\label{fig:dsig_dw10}
\end{figure}

A lower cut on photon transverse momentum $p_t > 1$ GeV is necessary to get rid of the
soft region where the VDM-Regge contribution dominates. 
In Ref.~\cite{KGSSz2016} we observed that the bigger distance between photons, 
the larger two-gluon to box contribution ratio is. 
Therefore we consider also a possibility to observe photons with forward calorimeters
(FCALs). In Fig. \ref{fig:dsig_dw10} we show differential distribution as a function of 
invariant mass of photons.
The results are shown both for box and two-gluon exchange mechanisms. For comparison
we also show cross section for the $\gamma \gamma \to e^+ e^-$ - subprocess. 
We emphasise that this subprocess is, however, 
a reducible background to the light-by-light scattering.

\begin{figure}
\centering
\includegraphics[scale=0.28]{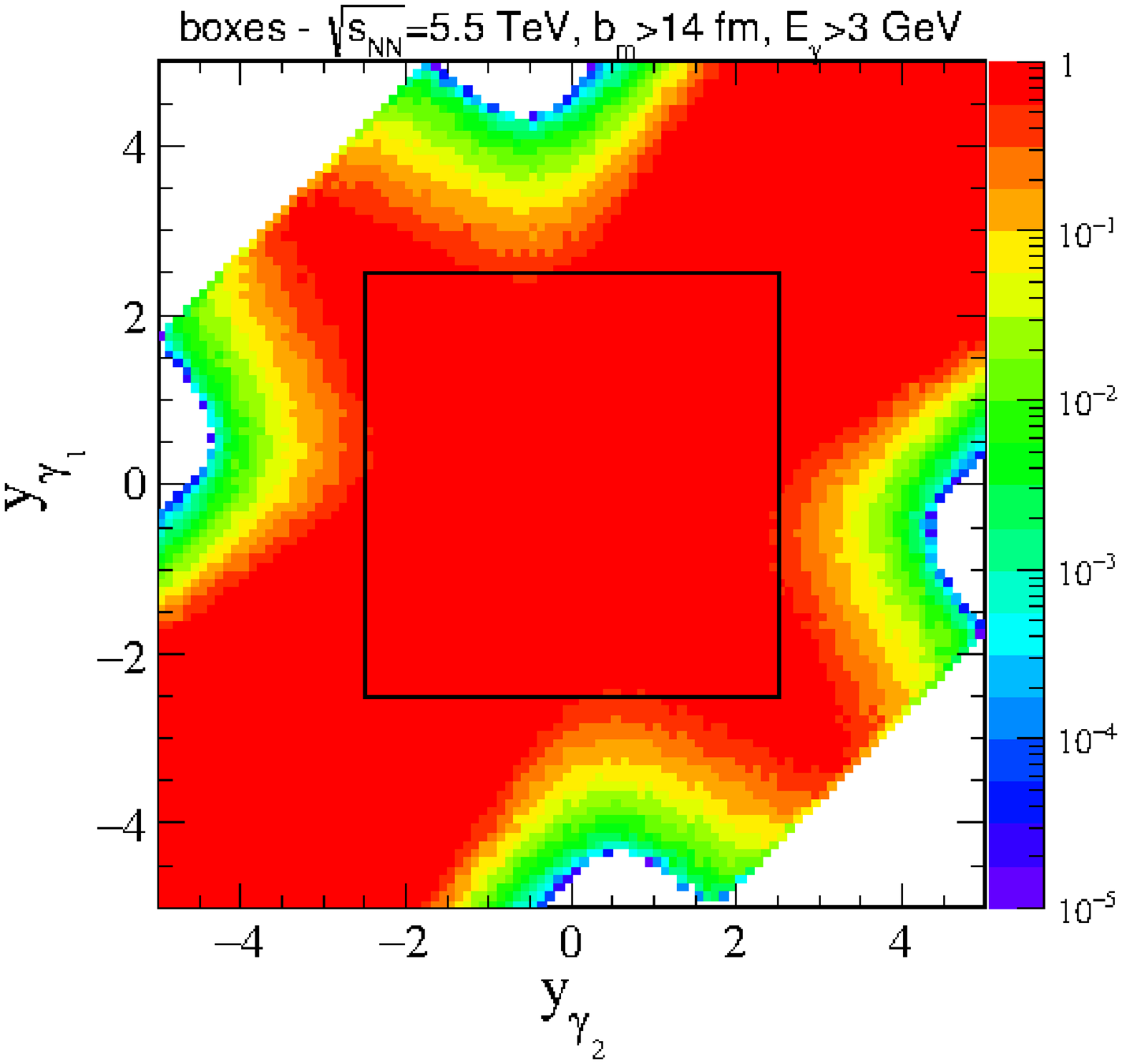}
\includegraphics[scale=0.28]{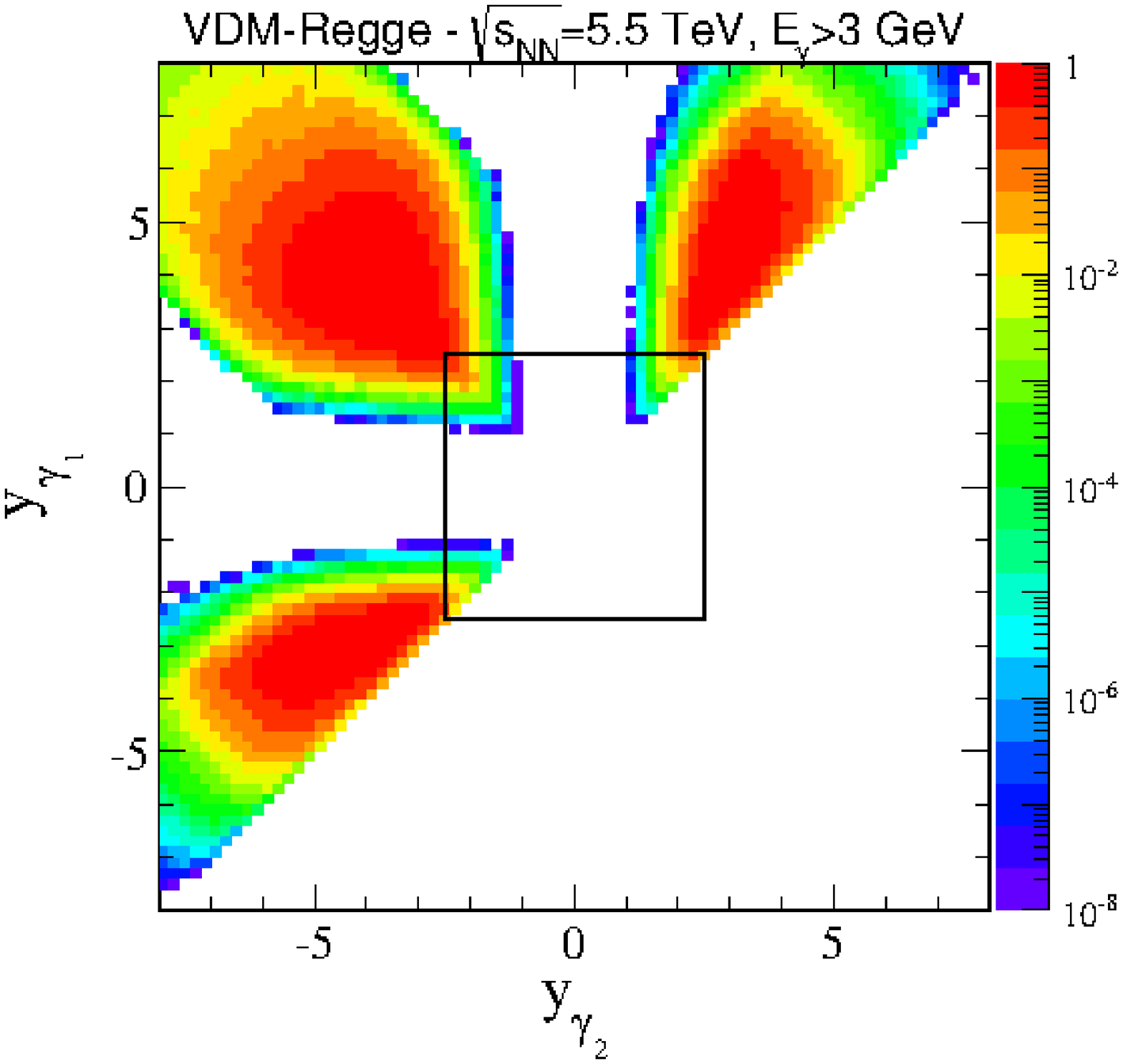}
\caption{\label{fig:dsig_dy1dy2}
Contour representation of two-dimensional 
($\mathrm{d} \sigma / \mathrm{d} y_{\gamma_1} \mathrm{d} y_{\gamma_2}$ in nb) 
distribution in rapidities 
of the two photons in the laboratory frame for box (left panel) 
and VDM-Regge (right panel) contributions. Nuclear calculations are done for $\sqrt{s_{NN}}= 5.5$ TeV.}
\end{figure}

If we try to answer the question whether the reaction can be measured with the help of LHC detectors
then we have to generalize Eq.~(\ref{eq:EPA_sigma_final_5int}) by adding extra integration over
additional parameter related to angular distribution for the subprocess \cite{KGLS2016}.
Fig.~\ref{fig:dsig_dy1dy2} shows two-dimensional distributions in photon rapidities 
in the contour representation. The calculation was done at the LHC energy
$\sqrt{s_{NN}}=5.5$ TeV. Here we impose cuts on energies of photons 
in the laboratory frame ($E_{\gamma}>3$ GeV). Very different distributions
are obtained for boxes (left panel) and VDM-Regge (right panel). 
In both cases the influence of the imposed cuts is significant.
In the case of the VDM-Regge
contribution we observe as if non continues behaviour 
which is caused by the strong transverse momentum dependence of the elementary cross section
(see Fig.~4 in Ref.~\cite{KGLS2016})
which causes that some regions in the two-dimensional space are 
almost not populated. Only one half of the ($y_{\gamma_1},y_{\gamma_2}$) space is shown for
the VDM-Regge contribution. The second half can be obtained from the symmetry
around the $y_{\gamma_1}=y_{\gamma_2}$ diagonal.
Clearly the VDM-Regge contribution does not fit
to the main detector ($-2.5<y_{\gamma_1},y_{\gamma_2}<2.5$) and extends towards large rapidities.
In the case of the VDM-Regge contribution we show much broader range of rapidity
than for the box component. We discover that maxima
of the cross section associated with the VDM-Regge mechanism are at
$|y_{\gamma_1}|,|y_{\gamma_2}| \approx$ 5. Unfortunately this is below the limitations 
of the ZDCs ($|\eta| > 8.3$ for ATLAS (\cite{ATLAS2007})
or $8.5$ for CMS (\cite{Grachov2008})).

\begin{figure}
\centering
\includegraphics[scale=0.28]{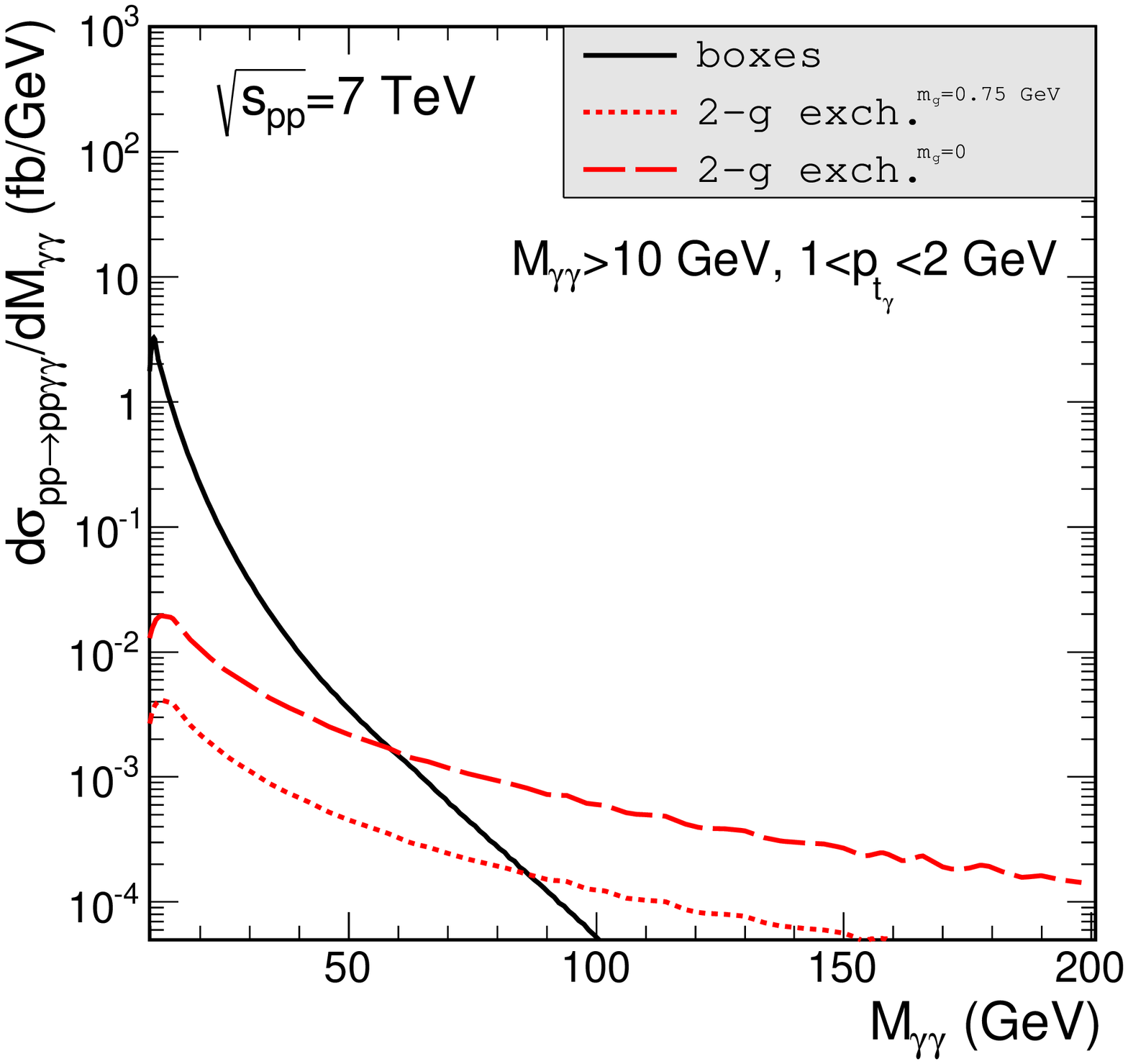}
\includegraphics[scale=0.28]{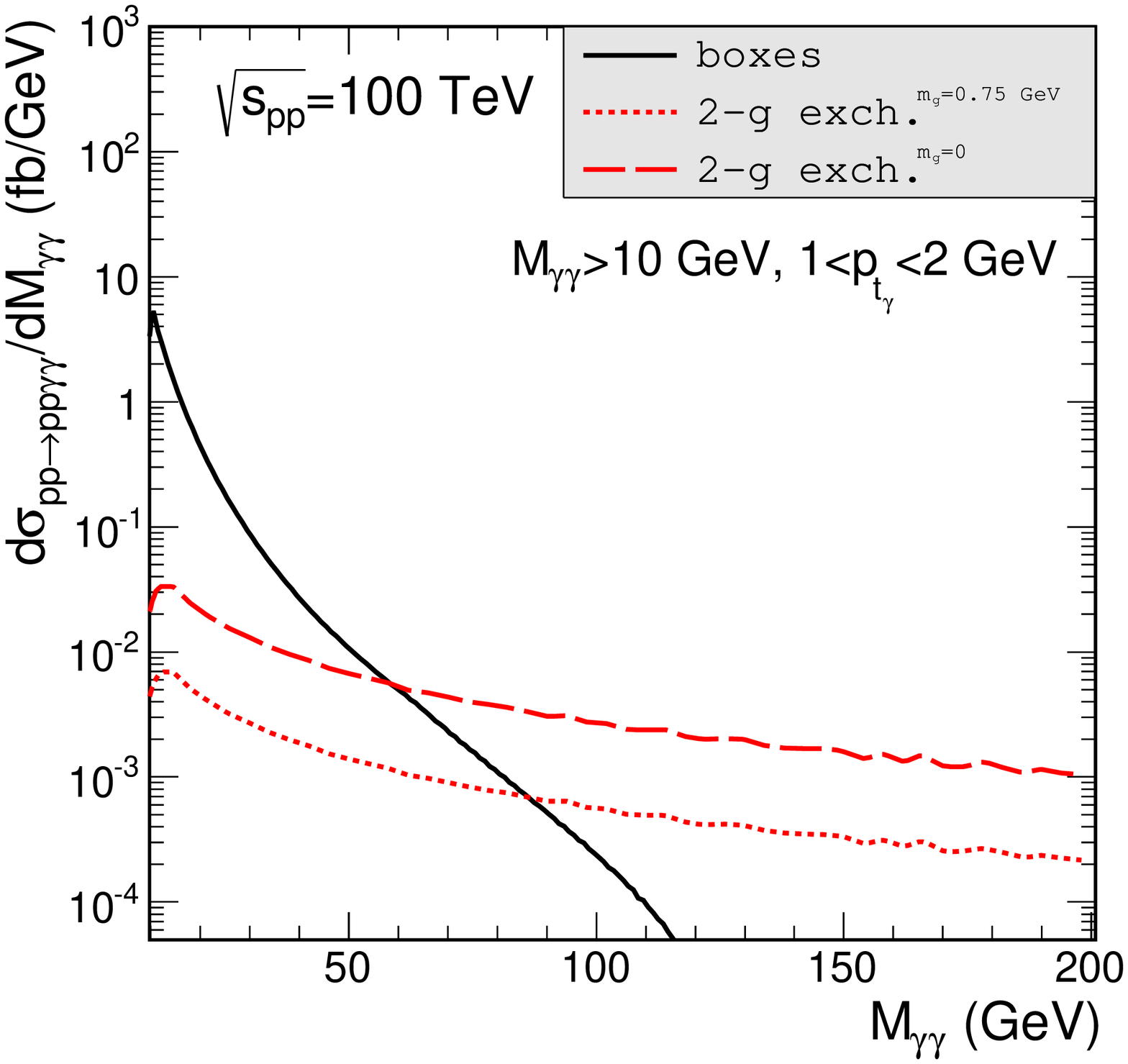}
\caption{\label{fig:dsig_dW_LHC_FCC}
Distribution in invariant mass of the produced photons for $\sqrt{s_{pp}} = 7$~TeV (LHC) 
and $\sqrt{s_{pp}} = 100$ TeV (FCC) for cuts on photon transverse momenta specified in the figure legend. No cuts on photon rapidities are applied here.}
\end{figure}

The nuclear distribution in the diphoton invariant mass is shown 
in Fig.~\ref{fig:dsig_dW_LHC_FCC}. 
The two-gluon distribution starts to dominante over the box contribution only above 
$M_{\gamma\gamma} > 50$ GeV for $1$ GeV $< p_t < 2$ GeV. 
However, the cross section in this region is rather small.
The situation for LHC (left panel) and for Future Circular Collider (FCC) energy (right panel) 
is rather similar. The dominance of the two-gluon exchange over the box contribution takes
place more or less at the same diphoton invariant masses.

\begin{figure}[!h]  
\center
\includegraphics[scale=0.28]{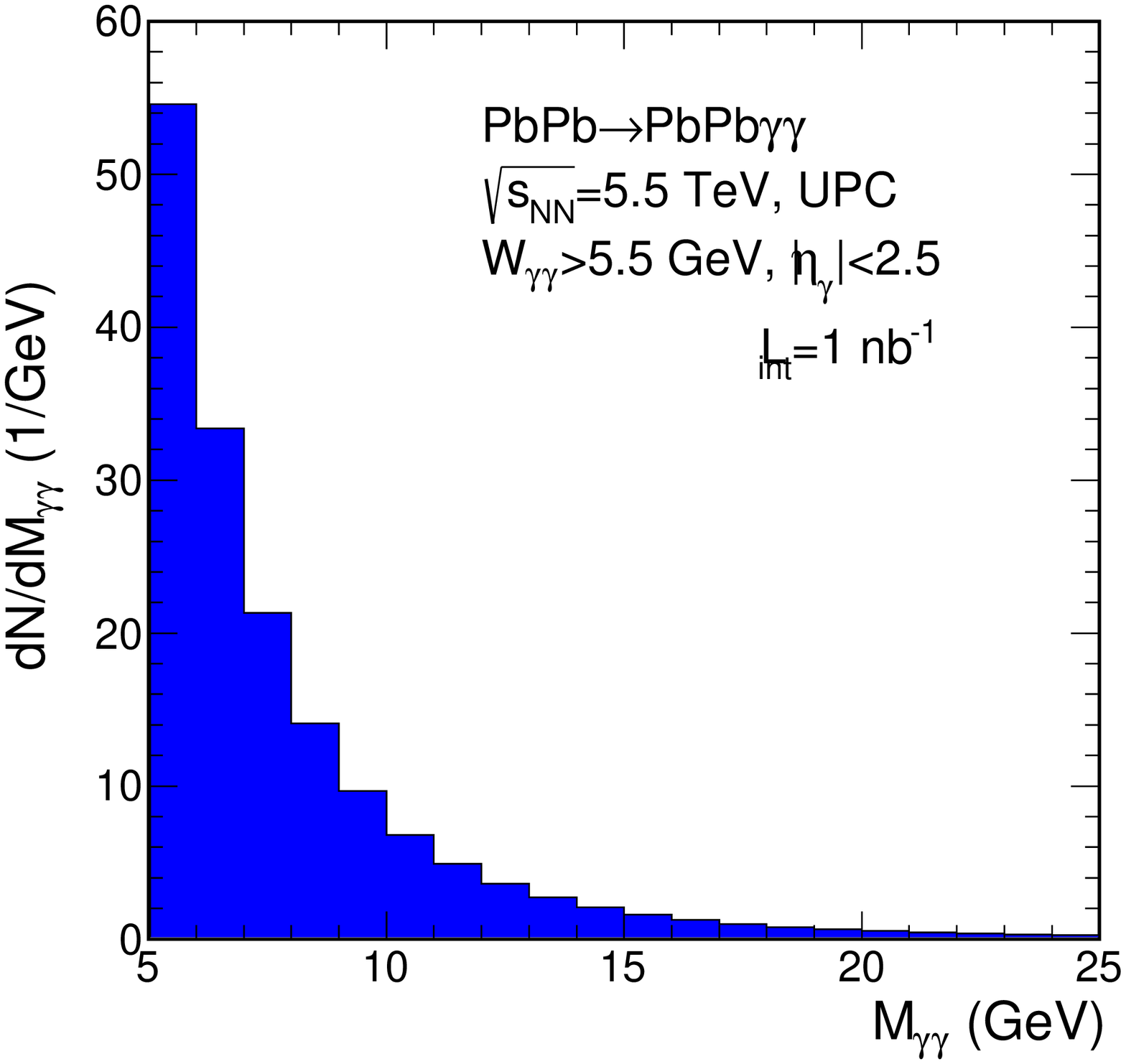}
  \caption{\label{fig:number_of_counts}
  Distribution of expected number of counts in $1$~GeV bins for cuts on 
  $W_{\gamma\gamma}>5.5$ GeV and $\eta_\gamma<2.5$. 
}
\end{figure}

Finally in Fig.~\ref{fig:number_of_counts} we show numbers of counts
in the $1$ GeV intervals expected for assumed integrated luminosity: $L_{int}=1$~nb$^{-1}$
typical for UPC at the LHC.
We have imposed cuts on photon-photon energy and
(pseudo)rapidities of both photons.
It looks that one can measure invariant mass distribution up to 
$M_{\gamma \gamma} \approx 15$ GeV.

\section{Summary}

In our recent papers \cite{KGLS2016,KGSSz2016} we have studied in detail 
how to measure elastic photon-photon scattering
in ultrarelativistic ultraperipheral lead-lead collisions 
and in $pp \to pp \gamma\gamma$ reaction.
The nuclear calculations were performed in an equivalent photon approximation 
in the impact parameter space.
The calculation for proton-proton collisions were done as usually in the
parton model with elastic photon distributions expressed in terms of proton electromagnetic
form factors.
The cross section for photon-photon scattering was calculated
taking into account well known box diagrams with elementary standard model particles
(leptons and quarks), a VDM-Regge component which was considered only recently \cite{KGLS2016}
in the context of $\gamma\gamma \to \gamma\gamma$ scattering
as well as a two-gluon exchange, 
including massive quarks, all helicity configurations of photons and massive and massless gluon.
Several distributions in different kinematical variables were calculated.
For $AA \to AA \gamma\gamma$ and $pp \to pp\gamma\gamma$ reactions
we identified regions of the phase space where the
two-gluon contribution should be enhanced relatively to the box contribution. The region
of large rapidity difference between the two emitted photons and intermediate transverse
momenta $1$ GeV $< p_t < 2-5$ GeV seems optimal in this respect.

Using the monopole form factor we get similar cross section to that found in \cite{d'Enterria:2013yra}
(after the correction given in Erratum of \cite{d'Enterria:2013yra}). 
Nevertheless, we think that application of realistic charge distribution 
in the nucleus gives more precise results.
We have shown an estimate of the counting
rate for expected integrated luminosity. We expect non-zero counts 
for subprocess energies smaller than $W_{\gamma \gamma} \approx$ 15-20 GeV. 

We have considered also an option to measure both photons by the forward calorimeters.
It is rather difficult to distinguish photons from electrons in FCALs. In heavy-ion collisions,
in addition, the cross section for $AA \to AA e^+e^-$ is huge, so this option seems not realistic.
In $pp \to pp \gamma\gamma$ case the corresponding background would be smaller 
but the signal is also reduced.

Recently, the ATLAS Collaboration published a note \cite{ATLAS_conf}
about evidence for light-by-light scattering signatures in quasi-real photon interactions 
from ultraperipheral lead-lead collisions at $\sqrt{s_{NN}}= 5.02$ TeV.
The data set was recorded in 2015 and corresponds to 480 $\mu$b.
The measured fiducial cross section which includes limitation on 
photon transverse momentum, photon pseudorapidity, diphoton invariant mass,
diphoton transverse momentum and diphoton acoplanarity,
has been measured to be $70 \pm 20$ (stat.) $\pm 17$ (syst.) nb, 
which is compatible with our predicted value of $49 \pm 10$ nb.


\begin{thebibliography}{99}


\bibitem{d'Enterria:2013yra}
D. d'Enterria and G. G. da Silveira, 
Phys. Rev. Lett. 111 (2013) 080405, 
Erratum: Phys. Rev. Lett. 116 (2016) 129901, 

\bibitem{KGLS2016}
M. K{\l}usek-Gawenda, P. Lebiedowicz and A. Szczurek,
Phys. Rev. \textbf{C93} (2016) 044907,

\bibitem{Bardin2009}
D. Bardin, L. Kalinovskaya and E. Uglov, 
Phys. Atom. Nucl. \textbf{73} (2010) 1878,

\bibitem{Lebiedowicz2013}
P. Lebiedowicz, R. Pasechnik and A. Szczurek, 
Nucl. Phys. \textbf{B881} (2014) 288,

\bibitem{Jikia1993}
G. Jikia and A. Tkabladze, 
Phys. Lett. \textbf{B323} (1994) 453,

\bibitem{Bern2001}
Z. Bern, A. De Freitas, L. J. Dixon, A. Ghinculov and H. L. Wong, 
J. High Energy Phys. 11 (2001) 031,

\bibitem{Gin87}
I. F. Ginzburg, S. L. Panfil and V. G. Serbo, 
Nucl. Phys. \textbf{B284} (1987) 685,

\bibitem{KGSSz2016}
M. K{\l}usek-Gawenda, W. Sch{\"a}fer and A. Szczurek, 
Phys. Lett. \textbf{B761} (2016) 399,

\bibitem{KG2010}
M. K{\l}usek-Gawenda and A. Szczurek,
Phys. Rev. \textbf{C82} (2010) 014904,

\bibitem{Dress}
M. Drees and D. Zeppenfeld, 
Phys. Rev. \textbf{D39} (1989) 2536,

\bibitem{ATLAS2007}
P. Jenni, M. Nessi and M. Nordberg,
Report No. LHCC-I-016, CERN-LHCC-2007-001,

\bibitem{Grachov2008}
O. A. Grachov et al. (CMS Collaboration),
J. Phys. Conf. Ser. 160 (2009) 012059,

\bibitem{ATLAS_conf}
The ATLAS collaboration, 
{\em Light-by-light scattering in ultra-peripheral Pb+Pb
collisions at $\sqrt{s_{NN}}=5.02$ TeV with the ATLAS
detector at the LHC} (ATLAS-CONF-2016-111).

\end{thebibliography}
\end{document}